\documentclass[12pt]{JHEP3}
\usepackage{amssymb,amsfonts}
\usepackage{showkeys}
\usepackage{amsmath}
\usepackage{fancybox}
\usepackage{frame}
\def\hri#1#2{\href{http://arxiv.org/abs/#1}{[ArXiv:#1]#2}}
\def\hre#1#2{\href{http://arxiv.org/abs/#1/#2}{[ArXiv:#1/#2]}}

\def\pa{\partial}

\relax

\def\be{\begin{equation}}
\def\ee{\end{equation}}
\def\bea{\begin{eqnarray}}
\def\eea{\end{eqnarray}}

\newcommand\fverb{\setbox\pippobox=\hbox\bgroup\verb}
\newcommand\fverbdo{\egroup\medskip\noindent%
                        \fbox{\unhbox\pippobox}\ }
\newcommand\fverbit{\egroup\item[\fbox{\unhbox\pippobox}]}

\newcommand{\bear}{\begin{eqnarray}}

\newcommand{\eear}{\end{eqnarray}}
\newbox\pippobox

\def\l{\lambda}

\def\e{\epsilon}

\def\sp{\;\;\;,\;\;\;}

\def\sq
\def\y{\psi}

\title{Asymptotic freedom, asymptotic flatness  and cosmology.}

\author{{\Large Elias ~Kiritsis$^{a,b}$}
~\\
~\\
$^a$\href{http://www.apc.univ-paris7.fr/APC_CS/}{APC, Universit\'e Paris 7}, CNRS/IN2P3, CEA/IRFU, Obs. de Paris, Sorbonne Paris Cit\'e, B\^atiment Condorcet, F-75205, Paris Cedex 13, France (UMR du CNRS 7164)
~\\
~\\
$^b$\href{http://hep.physics.uoc.gr/}{Crete Center for Theoretical Physics}, Department of Physics\\ University of Crete, PO Box 2208, 71003 Heraklion, Greece
}

\preprint{CCTP-2013-11}      

\abstract{Holographic RG flows in some cases are known to be related to cosmological solutions. In this paper another example of such correspondence is provided. Holographic RG flows giving rise  to asymptotically-free $\beta$-functions have been analyzed in connection with holographic models of QCD. They are shown upon Wick rotation to provide a large class of inflationary models with logarithmically-soft inflaton potentials.
The scalar spectral index is universal and depends only on the number of e-foldings.  The ratio of tensor to scalar power depends on the single extra real parameter that defines this class of models. The Starobinsky inflationary model as well as the recently proposed models of T-inflation are members of this class. The holographic setup gives a completely new (and contrasting) view to the stability, naturalness and other problems of such inflationary models.       }


\begin{document}

\maketitle 

\section{Introduction}

The recent Planck data, \cite{planck}, have constrained importantly models of  inflation, and gave prevalence to single field models with a very flat potential. A paragon of such models is the Starobinsky model \cite{staro}, but there are also other models in this class.
In fact the original Starobinsky model consisted of Einstein gravity coupled to a scale invariant theory (CFT), and took into account the conformal anomaly. It can be reformulated as an $R+R^2$ gravitational theory. This idea was generalized in \cite{brane-bulk} where the CFT was taken to have a large number of degrees of freedom (so that holography can be used) and was coupled to a non-conformal ``matter" theory with ${\cal O}(1)$ degrees of freedom.
The resulting class of models correspond via holography to a 4d brane (where matter lives) exchanging energy with a 5d bulk gravitational theory.

There has been discussion of such models of late, motivated by the Planck data, \cite{planck}. The conceptual setting and associated problems were discussed in \cite{stein}, generalizations have been discussed in \cite{kl} and the implementation of such models in supergravity in \cite{ellis,b,kehagias}.

The purpose of the present note is to draw some parallels between holographic models and inflationary models, and indicate  that from the holographic point
of view,the Starobinsky and related models belong to a larger class of models
that have marginal operators and asymptotic freedom, \cite{af} as their defining characteristics.
We will call these models Asymptotically-Flat(Free) Inflationary Models (AFIM), and they subsume the ones discussed in \cite{kl,ellis,kehagias}.

In AFIM,  the scalar spectral index is universal and depends only on the number of e-foldings.  The ratio of tensor to scalar power depends on the single extra real parameter that parametrizes this class of models.
Moreover, holographic analogies, shed a new light in the problems of the inflationary scenario, and suggest that AFIM may be more natural that hitherto suspected.

\section{Holography vs Cosmology}

Many researchers have noticed similarities between holographic saddle points with Poincar\'e symmetry, and cosmological solutions with maximal spatial symmetry.

In the most mild form,  the ``matter sector driving the inflation " can be described holographically (see \cite{brane-bulk} for an example).

One can however include also the gravitational sector in the story. In that case the two classes of gravitational solutions are related by a specific ``analytic continuation" that turns time into the holographic (spacelike) coordinate and vice versa. The proper way to perform this is to take (in 4d) $M_p^2\to -M_p^2$ together with $V\to -V$, \cite{skenderis}.
Under this map:
\begin{itemize}

\item deSitter space that describes an eternally inflating universe maps to Anti-de-Sitter space that describes a Conformally Invariant QFT (CFT), in one less dimension. The inverse of the Hubble constant, $H$, turns into the AdS curvature (length) scale $\ell$.

\item The asymptotically AdS boundary that corresponds to the UV region of the dual QFT, maps into a universe whose scale factor becomes asymptotically large.

\item An ``inflationary period" in cosmology corresponds to RG evolution in a scaling region driven by an almost marginal operator.

\item An inflating universe or increasing scale factor corresponds to an (inverse)  Renormalisation Group (RG) flow towards the UV. A shrinking universe corresponds to a conventional RG flow towards the IR.\footnote{In the context of Mirage Cosmology, \cite{mirage}, that acquires a holographic meaning when the bulk space is AdS-like, an expanding (probe-brane) universe corresponds to the brane moving away from the central black brane, while a shrinking universe corresponds to a brane approaching the central black brane.}

\item The ratio ${H^2\over M_p^2}$ that controls the size of inhomogeneities of inflationary perturbations, turns to $(M_p\ell)^{-2}$ that is proportional to $N^{-2}$ where roughly $N$ counts the degrees of freedom (``colors"), \cite{skenderis}. Therefore ${H^2\over M_p^2}\simeq 10^{-10}$ translates to $N\simeq 10^{3-4}$ (the ambiguity in the exponent depends on the $O(1)$ prefactor that may be different in different theories). This justifies a large-N limit.

    It is noteworthy that the hierarchy  ${H^2\over M_p^2}\simeq 10^{-10}$ seems technically unnatural/unstable  in gravity, \cite{stein}, but it is perfectly technically natural in the holographically dual QFT.

\item  An inflating universe corresponds to an  RG flow towards a UV fixed point (relevant operator), or away from an IR fixed point (irrelevant operator) . The period of inflation corresponds to the time the system spends in the  scaling region of the fixed point. The flow in the scaling region is controlled in the holographic dual by the scaling dimension of the operator dual to the inflaton scalar.

    \begin{itemize}

    \item A relevant operator corresponds to an exponentially evolving inflaton moving towards a minimum of the potential (UV fixed point).

\item    An irrelevant operator corresponds to an exponentially evolving inflaton moving away from a maximum of the potential (IR fixed point).

   \item  An intermediate case corresponds to marginally relevant or irrelevant operators. These give rise to a scaling region (=inflationary period) controlled by asymptotic freedom.

       \item Finally at the boundaries of field space (as the scalar runs off to infinity) there is a new type of scaling region, associated in holography with generalized criticality and hyperscaling violation, \cite{hyper}. In cosmology this corresponds to super-Planckian inflation which dominates many inflationary models.

       \end{itemize}

  \item There are two independent solutions to the second order equations for the scalar. In holography, near a UV fixed point the dominant solution is the source solution associated to the coupling of the relevant operator. The subdominant one is associated to the vev.
      In the scaling regime the vev solution is subleading.
      For IR fixed points the same applies to irrelevant operators.

      In cosmology, the analogous phenomenon is that the sub-leading solution dies out very fast.

There is however a crucial inversion between holography and cosmology.
Consider the RG flow between the (scaling) neighborhood of a UV fixed point towards an IR fixed point driven by a relevant scalar operator of UV scaling dimension\footnote{$d$ is the dimension of the boundary UV CFT.}
 $\Delta_{UV}<d$ and IR scaling dimension $\Delta_{IR}>d$.
For the gravitational dual theory, in the UV scaling region, the two linearly independent solutions, the dominant (source), and the subdominant (vev) vanish at the UV fixed point.
In the IR, one of the solutions blows up , while the other is regular.

The correct regular solution for the flow, involves a special combination (tuning)  of the two solutions in the UV scaling region so that they asymptote to the regular solution in the IR fixed point.
Therefore the subdominant solution in the neighborhood of the UV fixed point starts growing during the RG flow, and when it becomes of the order of the dominant solution the scaling regime ends (corresponding to the end of inflation in a cosmological setup).
On the other hand in the scaling regime of the IR fixed point, only the subdominant solution is present due to the UV tuning for regularity.

In holography this flow corresponds to the flow in a scalar potential from the region of a maximum to the minimum of the potential.

Consider the analogous time evolution in the cosmological context from the neighborhood of a maximum of the potential to a minimum. Near the maximum, there are two solutions for the scalar, one regular while the other is singular. Tuning of the initial conditions is needed, if we would like our solution to be valid arbitrarily far in the past.
Once we arrive in the scaling region of the minimum, both solutions are regular and asymptote to the relevant value at the minimum.

  Therefore in holography regularity is imposed at the end-point of the flow, while in cosmology at  the starting (fixed) point.

\item The scalar and tensor cosmological perturbations correspond to the scalar and tensor fluctuations in holography, that determine the two-point functions of the scalar operator and the stress tensor in the scaling regime.

\item An exit from inflation corresponds potentially to one of two effects. The first is that the subleading solution picks up and destroys slow roll. In the holographically dual description this corresponds to leaving the scaling region. The other possibility is that another field intervenes, so that the inflaton exits inflation. In the holographically dual description it corresponds to another operator becoming relevant (turned-on and back-reacting on the flow) so that the flow misses the fixed point in question.

\item     In fact, a universe that starts in the infinite past as near dS, passes an intermediate region where it inflates, then exits inflation, and then at future infinity re-inflates again is very similar to a RG flow between two fixed point CFTs, while the flow in the middle, remains close to a third fixed point, for at least 60 e-foldings. Holographic models of that type correspond to walking QFTs and a large class can be found in \cite{jk}, some of them being very similar  to QCD in the Veneziano limit.

\item In holographic models in \cite{jk}, the avoidance of the intermediate fixed point is triggered by the fact that the scalar operator that drives the flow has complex anomalous dimension in the fixed point in question, and the scalar violates the BF bound.
    In the cosmological incarnation, exit from inflation is traced to the mass of the inflaton becoming large enough.

\end{itemize}

It is not clear whether the cosmology/QFT  correspondence painted above has any deep structure beyond the similarities at the level of semiclassical solutions.
We will take it however at face value and we will explore its consequences.
In particular, we will focus on a conformal (asymptotically AdS) scaling regime that is driven by an marginally relevant  operator,  a case made famous from asymptotically  free gauge theories.

\section{Holography and asymptotic freedom.}

We will use a single scalar, dual to a scalar operator, and the effective holographic action  after field redefinitions will be
\be
\label{1}
S_{d+1} =M^{d-1}\int d^{d+1}x \sqrt{-g}
\biggl\{R -{1\over 2} \pa_{\mu}\phi\pa^{\mu}\phi+V(\phi)\biggl\}
\ee
Note that we use the opposite to the standard sign for the potential, as we are in the AdS regime (where $V>0$).

We will  use a holographic  coordinate $u$. The boundary region of an asymptotically $AdS$ solution now lies at $u = + \infty$. Using these coordinates, the ansatz for the metric takes the form:
\be
ds^2 = du^2 + e^{2A(u)} dx_{\mu} dx^{\mu}.
\label{2}\ee
and the equations of motion are
\be
d(d-1)\dot{A}^2-{1\over 2}\dot\phi^2 - V(\phi) = 0
\label{7}\ee
\be
2(d-1) \ddot{A} + \dot{\phi}^2 = 0
\label{8}\sp
\ddot{\phi} + d \dot{A} \dot{\phi} + { dV(\phi) \over d \phi}=0.
\ee
where $\dot{}$ stands for derivative with respect to the holographic coordinate $u$.

The radial coordinate is interpreted as a holographic RG scale and the flow in $u$ as the holographic RG flow.
In QFT a standard concept that controls the RG flow is the $\beta$ function. This can be defined by introducing the ``superpotential" $W(\phi)$ as a solution of the non-linear equation
\be
\label{10}
{dW^2\over 2(d-1)}-W'^2=2V(\phi).
\ee
Then the equations
\be
\label{11}
\dot A = -{W\over 2(d-1)} \sp \dot\phi= {dW\over d\phi}\equiv W' \;,
\ee
are equivalent to the equations of motion in (\ref{7},\ref{8})\footnote{Note that there is a one parameter family of superpotentials $W$ solving (\ref{10}) that is important to match all parameters the solutions depend upon. The topology of the space of solutions can be however complicated, \cite{t}. Most of the solutions for $W$ give rise to singular gravitational solutions that are not acceptable.}.

The $AdS_{d+1}$ solution is:
\be
A(u) = { u\over \ell} \; , \; \phi(t) = \phi_*.
\label{3}\ee
with $\phi_*$ an extremum of the potential, $V'(\phi_*)=0$ and $V(\phi_*)={d(d-1)\over \ell^2}$.
Depending whether $V''(\phi_*)$ is positive or negative we have a UV or an IR fixed point of the RG flow.

\subsection{The 5d holographic case}

To be concrete, we choose $d=5$ as this is holographically dual to 4d QFTs.
A similar analysis however, with the same conclusions can be done in any $d>2$.

In this case the superpotential  $W(\phi)$ satisfies:
\be
\label{4}
{2W^2\over 3}-W'^2=2V(\phi).
\ee
where $'$ is derivative with respect to $\phi$ and the first order equations (\ref{11}) become:
\be
\label{5}
\dot A = -{W\over 6} \sp \dot\phi= {dW\over d\phi}\equiv W' \;,
\ee

We define
\be
\l=e^{a\phi}
\label{19}\ee
This combination is typically a coupling constant in string theory. Examples includes the string coupling constant $g_s=e^{\phi}$ where $\phi$ is the string dilaton, radii of compactifications, as well as volumes of compact cycles in CY manifolds. They all have logarithmic kinetic terms that prompts the interpretation above.

We also consider the regime in the limit $a\phi\to -\infty$, so that $\lambda\to 0$.
In this limit, an asymptotically AdS potential can be expanded as
\be
V(\phi)\simeq {12\over \ell^2}\left[1+\sum_{n=1}^{\infty}V_n~\l^n\right]
\label{6}\ee
Note that by shifting $\phi$ by a constant, $\phi\to\phi+\phi_0$, the first coefficient $V_1$ can be changed at will.
The potential is chosen positive so that the AdS point is a minimum corresponding to a UV CFT. Then all coefficients $V_n\geq 0$.

Also note that the AdS minimum, has the very special property that
\be
{d^n V\over d\phi^n}\Big |_{\l\to 0}=0~~~~\forall n\in N
\label{20}\ee
We will call these potentials Asymptotically Flat (AF), and as we will soon show they give rise (deservedly) to Asymptotically Free (AF) flows.
In this sense these are the shallowest possible AdS minima as the dual operator is marginal.

We can solve, perturbatively (\ref{4}) for the superpotential, in the scaling regime $\l\to 0$.
\be
W=-{6\over \ell}\left[1+\sum_{n=1}^{\infty}W_n~\l^n\right]\sp W_1={V_1\over 2}\sp W_2={V_2\over 2}-\left(1-{3\over 2}a^2\right){V_1^2\over 8},
\label{21}\ee
and so on. The sign of $W$ was chosen so that it leads to an AdS boundary.

We may now define the conventional $\beta$-function as
\be
{d\l\over dA}=a\l{d\phi\over dA}=-6a\l{W'\over W}=-b_0\l^2-b_1\l^3+\cdots\equiv \beta (\lambda)
\label{22}\ee
keeping in mind that in holography, in the asymptotically-AdS regime, $A$ represents the logarithm of the RG scale.
Matching (\ref{22}) with the solution one obtains
\be
b_0=6a^2W_1=3a^2V_1\sp b_1=6a^2(2W_2-W_1^2)=6a^2\left[V_2-{V_1^2\over 2}\left(1-{3\over 4}a^2\right)\right],
\label{23}\ee
and so on.
Solving (\ref{5}) we obtain the first order corrections to the AdS solution
\be
\lambda\simeq {1\over b_0 A}\simeq {\ell\over 3a^2V_1 u}+\left[{2-3a^2\over 36 a^4}-{4\over 9a^4}{V_2\over V_1^2}\right]{\log u\over V_1u^2}
\label{24} \ee
 \be
 A={u\over \ell}+{1\over 6a^2}\log u+\left[{-2+3a^2\over 144 a^4}+{1\over 18a^4}{V_2\over V_1^2}\right]{\ell\over u}
+\left[{-2+3a^2\over 72 a^4}+{1\over 9a^4}{V_2\over V_1^2}\right]{\ell\log u\over u}
\label{25}\ee

The scaling symmetry implies there is a natural redefinition
\be
\hat \lambda \equiv V_1 \l
\label{26}\ee
so that the equations above become
\be
{d\hat \l\over dA}=\beta(\hat\l)=-b_0\hat\l^2-b_1\hat\l^3+\cdots
\label{27}\ee
\be
b_0=3a^2\sp b_1=6a^2\left[{V_2\over V_1^2}-{1\over 2}\left(1-{3\over 4}a^2\right)\right]
\label{28}\ee
\be
\hat \lambda\simeq {\ell\over 3a^2 u}+\left[{2-3a^2\over 36 a^4}-{4\over 9a^4}{V_2\over V_1^2}\right]{\log u\over u^2}
\label{29} \ee

We summarize this section as follows:

\begin{itemize}

\item Asymptotically Flat potential minima, as in (\ref{20}),  correspond (holographically) in a one-to-one correspondence to asymptotically free flows, as captured by the holographic $\beta$-function in (\ref{22}).

\item The Asymptotically flat regime of the potential corresponds to the perturbativity of the $\beta$-function.

\item The associated operator is nearly marginal (marginality breaks at one loop). This is to be contrasted with operators whose scaling dimension is $\Delta=4-\eta$ with constant $\eta<<1$, which are slightly relevant at tree level.

\item The most important parameter of this class of potentials and flows is captured by the one-loop answer and is the real number $|a|$.
\item Such asymptotic potentials have been used to model the high-energy region of Improved Holographic QCD, \cite{ihqcd}, a semi-phenomenological holographic model for large-N YM (non-supersymmetric) theory in four dimensions. Although the asymptotically-free region is not reliable holographically, it provides the correct boundary conditions in such holographic models, in order to describe the IR physics reliably, \cite{data}.

  \item All such regimes/models have an approximate scaling symmetry\footnote{An alternative take on this approximate symmetry was suggested in \cite{kl}.}. It is violated by the same UV scaling violating logs found in QCD.
      A scaling transformation can be implemented by using the $\beta$- function.

\item From the point of view of gravity, it would seem that this situation is highly fine-tuned, as was argued recently in \cite{stein} in the cosmological context. However, from the point of view of the holographically dual theory, for example large N pure YM, it is not fine tuned at all. The theory naturally contains a single scalar (nearly) marginal operator ($Tr[F^2]$) which is dual to $\phi$ and NO relevant operators.\footnote{There is another marginal scalar operator that is a pseudoscalar, namely the instanton density $Tr[F\wedge F]$. This is exactly marginal in perturbation theory, but seems to become irrelevant non-perturbatively, \cite{ihqcd}.}

    It is also amusing that from the dual perspective, supersymmetry makes things worse. Supersymmetric theories have necessarily relevant operators, and the instabilities of \cite{stein} become relevant. The more the supersymmetry, the more relevant operators seem to appear, as the $N=4$ supersymmetric case in four dimensions seems to indicate.

\item An interesting  question is  whether marginally relevant operators can have potentials other than the AF ones in (\ref{20}). In string theory the answer seems to be negative, as this would require the potential to have a single power of the scalar greater than quadratic, \cite{bk}, and this is unlikely. On the other hand AF potentials van be found  in string theory, \cite{qu} and supergravity, \cite{kl,ellis,b,kehagias}.

\item ``Non-perturbative" terms can appear in the potential in the form of series in non-commensurate powers of $\lambda$. In fact generic string theory and gauged supergravity potentials contains multiple exponentials. However there are always scaling regimes where a single exponential dominates and the analysis above is reliable. Of course, depending on the nature of such extra terms, the regime of validity of AF potentials may be affected. However as the YM case indicates, this is not necessary.

\item The presence of other operators dual to scalars changes somewhat the picture above. In the presence of several marginally relevant operators, the story above is more or less unchanged. However, if we are in the UV scaling region, relevant operators are generically important. They are dual to ``tachyonic scalars".
     Even if their initial condition sets them to zero (source) the subleading solution may be force to be non-zero for regularity, and then they will dominate the flow (chiral symmetry breaking in QCD is in this class where the relevant scalar operator is the quark mass term).

There are also cases where once the leading solution is tuned to zero, the subleading solution remains zero. Even in that case however fine tuning is needed in the presence of of a relevant scalar operator. Similar remarks apply to the scaling region of  IR fixed points.

\end{itemize}

\section{Back to cosmology}

The action is the same as in (\ref{1}) with only $V\to -V$ but the metric ansatz is now a cosmological one
\be
ds^2 =-dt^2 + e^{2A(t)} dx_{i} dx^{i}.
\label{12}\ee

The cosmological equations are\footnote{To match them to standard convention in cosmology one should choose $M_p^2=2$.}
\be
d(d-1)\dot{A}^2-{1\over 2}\dot\phi^2 -V(\phi) = 0
\label{13}\ee
\be
2(d-1) \ddot{A} + \dot{\phi}^2 = 0
\label{14}\sp
\ddot{\phi} + d \dot{A} \dot{\phi} + { dV(\phi) \over d \phi}=0.
\ee

The superpotential equations remain the same
\be
\label{16}
{dW^2\over 2(d-1)}-W'^2=2V(\phi)\sp
\dot A = -{W\over 2(d-1)} \sp \dot\phi= {dW\over d\phi}\equiv W' \;,
\ee

The $dS_{d+1}$ solution is now:
\be
A(t) = { t \over \ell}\equiv Ht \; , \; \phi(t) = \phi_*.
\label{18}\ee
with $\phi_*$ an extremum of the potential, $V'(\phi_*)=0$ and $V(\phi_*)={d(d-1)H^2}$.

\subsection{4d cosmology}

We now pick the potential with the opposite overall sign and the dS point to be a maximum but otherwise it is still an AF potential.
\be
V(\phi)\simeq {6 H^2}\left[1+\sum_{n=1}^{\infty}V_n~\l^n\right]\sp \l\equiv e^{a\phi}
\label{6e}\ee
The equations are now
\be
\label{4e}
{3W^2\over 4}-W'^2=2V(\phi)\sp \dot A \equiv H(\phi)= -{W\over 4} \sp \dot\phi= {dW\over d\phi}\equiv W'
\ee
where $'$ is derivative with respect to $\phi$ and $H(\phi)$ is the Hubble scale. (\ref{4e}) indicates that $W$ is essentially the Hubble scale.

The solution is
\be
W=4H\left[1+{V_1\over 2}\l+\left[{4a^2-3\over 24}V_1^2+{V_2\over 2}\right]\l^2+\cdots\right]
\label{32}\ee
\be
\l=-{1\over 2a^2HV_1 t}-\left[{4a^2-3\over 24 a^4}+{V_2\over 2a^4 V_1^2}\right]{\log t\over V_1H^2 t^2}+\cdots
\label{36}\ee
\be
A=-Ht+{1\over 4a^2\log t}+\left[\left({4a^2-3\over 96 a^4 }+{V_2\over 8 a^4 V_1^2}\right)+ \left({4 a^2-3\over 48a^4} + { V_2\over 4 a^4  V_1^2}\right)\log t\right]{1\over Ht}+\cdots
\label{37}\ee

Using the standard formulae of slow-roll inflation we obtain
\be
\e={V'^2\over V^2}\simeq a^2(V_1\l)^2+2a^2\left[2{V_2\over V_1^2}-1\right](V_1\l)^3+\cdots
\label{38}\ee
\be
\eta=2{V''\over V}\simeq 2a^2V_1\l-2a^2\left[4{V_2\over V_1^2}+1\right](V_1\l)^2+\cdots
\label{30}\ee
There are two related functions
\be
\epsilon_H\equiv 4\left({H'\over H}\right)^2=4\left({W'\over W}\right)^2\sp \eta_H\equiv 4{H''\over H}= 4{W''\over W}
\label{39}\ee
When they are small, they signal slow roll inflation and in that limit they are given by $\e_H\simeq \e$, $\eta_H\simeq \eta-\epsilon$. Note that $\epsilon_H$ is related directly to the holographic $\beta$-function,   $\epsilon_H\sim {\beta^2\over \lambda^2}$.

The number of e-folds is given by
\be
N={1\over 2}\int^{\l_i}_{\l_e}{Vd\phi\over V'}={1\over 2}\int^{\l_i}_{\l_e}{V\over {dV\over d\l}}{d\l\over a^2\l^2}\simeq {1\over 2a^2 V_1 \l_e}-{1\over 2a^2 V_1 \l_i}
\label{31}\ee

We may therefore rewrite
\be
\e\simeq {1\over 4a^2N^2}\sp \eta\simeq -{1\over N}
\label{33}\ee

The spectral indices are
\be
n_s-1=-6\e+2\eta\simeq -{2\over N}+{\cal O}(N^{-2})  \sp n_{\rm grav}\simeq -2\e\simeq -{1\over 2a^2N^2}+{\cal O}(N^{-3})
\label{40}\ee
and  the tensor ratio
\be
 r\simeq ~12.4\e\simeq {3.1\over a^2N^2}+{\cal O}(N^{-3})
\label{34}\ee
The Planck constraint on $r$ suggests that $|a|\gtrsim 0.9$.
The running of the spectral index is given by
\be
a_s\equiv {dn_s\over d\log k}=-16\e\eta+24\e^2+2\xi^2\;.
\ee
Using
\be
 \xi^2=4{V'V'''\over V^2}\simeq 4a^4(V_1\l)^2+\cdots\simeq {1\over N^2}+\cdots
\ee
we obtain
\be
a_s={2\over N^2}+{\cal O}(N^{-3})
\ee

We may contrast this to the case where the inflaton is dual to a slightly relevant operator, \cite{bg}.
In that case, taking the fixed point to be at $\phi=0$.
\be
V={6H_0^2}\left[1+{\Delta(3-\Delta)\over 12}{\phi^2} +{\cal O}(\phi^3)\right]
\label{41}\ee
Solving perturbatively (\ref{4}) we obtain for the superpotential,
\be
W=-4H_0\left[1+{3-\Delta\over 8}\phi^2+{\cal O}(\phi^3)\right]
\label{42}\ee
and therefore
\be
\e_{H}={(3-\Delta)^2\over 4}\phi^2+{\cal O}(\phi^3)\sp \eta_H={3-\Delta}-{(3-\Delta)^2\over 16}\phi^2+{\cal O}(\phi^3)
\label{43}\ee
Even at $\phi=0$, $\eta_H$ is non-zero and therefore, $\Delta=3-\eta_H$ with $\eta_H\ll 1$ as claimed. The spectral index remains fixed as the number of e-foldings increases without bound.

It should be mentioned that models that belong to this class also include standard polynomial potentials (with $\phi^4$ as the maximum power, and and curvature dependent mass term, \cite{xi}. 

In the case where several scalars are present with a potential that is AF, most of the story above remains true. The scalar spectral index is still universal, and there is a slight change  in $r$ that reflects the presence of extra marginally relevant dual operators. However knowledge of these two parameters alone is not enough to tell the difference between one or many inflatons of that type.

\section{Conclusions and outlook}

In this note, after developing a map between (homogeneous) holographic and cosmological models we have focused and characterized a class of theories that on the holographic side correspond to the scaling regions of CFTs with one (or more) marginally-relevant (or marginally-irrelevant) operators.

In the holographic context, they are controlling flows that are asymptotically free. In the cosmological context they are Asymptotically Flat Inflationary Models controlled by a single real parameter in their inflating regions.
The spectral index and its variation  are universal for this class and depend on the number of e-foldings. The tensor to scalar density ratio $r$ depends also on the extra real parameter that characterizes this class of theories.

This universality is not unlike Wilson's universality of RG flow, but has the extra simplifying characteristics of marginally-relevant operators that make the characteristics of the flow even more universal.

In holographic asymptotically-free theories, there are three prominent scales that enter the scaling regime: the Planck scale $M_p$, the AdS curvature length $\ell$, and the scale $\Lambda$ generated by dimensional transmutation due to asymptotic freedom. In the dual physics only one basic scale appears. Therefore, theories like YM that are in this class, have no adjustable parameters, but only a scale.  From the dual point of view, $M_p\ell$ is controlling N, the number of colors while $\ell$ sets the scale. However, as was pointed out in \cite{data}, in the presence of marginally relevant flows the scale $\ell$ becomes unobservable as it can be changed using the scale covariance of the scaling region.

The translation of this in the cosmological setting is that although two distinct scales are expected in inflationary histories in this class (namely the Hubble scale and the parameter that controls the power evolution of the scalar) again only one is independent as one can be scaled out of the problem.
This is the reason for the extra universality of AFIM.

The holographic point of view has the bonus of offering potentials explanations to several cosmological puzzles:

\begin{itemize}

\item Fine-tuning in the gravitational context is one of the hallmarks of inflation. From the point of view of gravity, it would seem that AFIM are even more  fine-tuned that generic inflationary  models,\cite{stein}.
    The notion of fine-tuning in the dual QFT is related to the presence of relevant operators in the theory. Theories that lack relevant operators but have only marginally relevant ones, are rare but exist. YM is the perfect example. Although a choice needs to be made, the choice is technically natural in the conventional sense of the word.

\item The issue that $H/M_p$ is very small, has been always considered as a major problem. The holographic thinking turns this on its head. It is as natural as choosing the number of colors $N$ in a gauge theory, because it is $N$ that controls this particular ratio of scales in holography.

 \item  From the dual perspective, supersymmetry makes the fine-tuning  more ``unstable". Supersymmetric theories have necessarily relevant operators, and the instabilities discussed in \cite{stein} become relevant. The more the supersymmetry, the more relevant operators seem to appear, as the $N=4$ supersymmetric case in four dimensions seems to indicate.

\item  There is an infinity of irrelevant operators in a CFT. From a holographic point of view, this suggests that there is an infinite-dimensional space of paths (in QFT space) that we can arrive in the IR to a given CFT. That same CFT has always (at least in $d>2$) a finite number of relevant operators, and therefore there is a finite-dimensional space  of paths leading flows out of that CFT to other CFTs in the IR.

    In holographic flows we must tune all irrelevant operators as we arrive in an IR fixed point in order to preserve the regularity of flow, that as we understand today, is a crucial ingredient for its admissibility.
Usually, this is accomplished in practice, by setting all scalars corresponding to UV irrelevant operators to zero, and tuning only the solution of the scalars that start relevant in the UV and necessarily turn irrelevant in the IR.

A similar statement can be made in cosmological flows. The only difference is that for an expanding universe regularity is non-trivial at early  times. Therefore, absence of the big-bang-singularity in this context (it is regularised by an initial dS space) involves the tuning of an infinite number of scalars (the would be dual to all irrelevant operators and appear as stringy states in a string theory description).

However as in holography, this tuning is necessary for regularity, or to put it differently, for the existence of a semiclassical space-time.

Interestingly enough, this infinite number of tunings looks natural in the context of holography, but is anathema in cosmology.
Holography suggests that it could be  the price we have to pay for being in a semiclassical state of the theory. There are good reasons to believe that most states, even at large $N$, are not semiclassical and cannot be described by smooth geometries. Therefore the question translates into why the geometry around us is semiclassical and not fully quantum. We can imagine that it is not always so, during the whole history of the universe, and holographic inflationary models of this type have been proposed, \cite{skenderis}. Another answer to that question could be, that this is because we are simply here.

\item It is intriguing that cosmological  data suggest two periods of acceleration one in the past (inflation) and one today (dark energy).
    The map to holography identifies this state of affairs with the simplest RG flow, from one UV fixed point to an IR one. The UV part is nearly marginal as data indicate, but we know much less about the IR part.

The map to RG flows also resolves a major puzzle of inflation: its presence in the past has always been thought to involve some form of fine tuning.
From the The RG point of view it is a necessity. An RG flow in the ultimate IR (this corresponds for expanding universes to the infinite past) {\em must} end up in a fixed point with no more relevant operators other-wise it will be unstable. Few CFTs, if any have this property, and none is known in in three or four dimensions.\footnote{They are extremely rare in 2d, but such examples are known, \cite{s}.}

In the absence of such rare stable CFTs, those having only marginally relevant operators come next, and this is precisely the case discussed here.

The RG arguments translated in a cosmological setting imply that such inflationary periods are natural and not fine tuned.

\item The minimal picture in the item above is not the only one tenable. The Universe may ``pass-by" one or more nearly scale invariant, scaling regions before ending up in the final accelerating regime.

\end{itemize}

In summary, translated in holographic language, AFIM, are as simple and as natural as $SU(N_c)$ YM in 4-dimensions, in the asymptotically free regime, with $N_c\sim 10^{3-4}$.

We have not solved any important problems in this note, but we think that the remarks concerning the interplay of holography and cosmology have an impact on some central problems in cosmology, are intriguing  and deserve being made public. Whether they will help guides us to the solution, remains to be seen.

\section{Acknowledgements}\label{ACKNOWL}

I would like to thank R. Kalosh,  A. Linde and K. Skenderis for discussions and correspondence.

Thanks also to S. J. Rey of Seoul National University and APCTP  for hospitality during the completion of this work.

This work was supported in part by grants PERG07-GA-2010-268246, PIF-GA-2011-300984, the EU program ``Thales'' and "HERAKLEITOS II'' ESF/NSRF 2007-2013 and was also co-financed by the European Union (European Social Fund, ESF) and Greek national funds through the Operational Program ``Education and Lifelong Learning'' of the National Strategic Reference Framework (NSRF) under ``Funding of proposals that have received a positive evaluation in the 3rd and 4th Call of ERC Grant Schemes''.



\end{document}